\documentclass [10pt] {revtex4}
\usepackage {amsmath}
\usepackage {amstext}
\usepackage {latexsym}
\usepackage {xspace}
\begin{document}
%\draft
%\preprint{HEP/123-qed}
% --- /home/prasad/.lib/tex/vgarticle.sty ---
% =====================================================================
% Personal article style			v guruprasad,sep1999
% =====================================================================

% ---------------------------------------------------------------------
% General favourite definitions

% Orderly dates 
\newcommand\DateYMD{
	\renewcommand\today{\number\year.\number\month.\number\day}
	}

\newcommand\DateDMY{
	\renewcommand\today{\number\day.\number\month.\number\year}
	}

\newcommand\DateDmmmY{
	\renewcommand\today{
		\number\day\space
		\ifcase\month\or
		Jan\or Feb\or Mar\or Apr\or May\or Jun\or
		Jul\or Aug\or Sep\or Oct\or Nov\or Dec\fi
		\number\year
		}
	}

% Tables:
\newenvironment{mptbl}{\begin{center}}{\end{center}}
\newenvironment{minipagetbl}[1]
	{\begin{center}\begin{minipage}{#1}
		\renewcommand{\footnoterule}{} \begin{mptbl}}%
	{\vspace{-.1in} \end{mptbl} \end{minipage} \end{center}}

% Figures:
\newif\iffigavailable
	\def\figavailable{\figavailabletrue}
	\def\nofigavailable{\figavailablefalse}
	\figavailable% default

\newcommand{\Fig}[4][t]{
	\begin {figure} [#1]
		\centering\leavevmode
		\iffigavailable\epsfbox {\figdir /#2.eps}\fi
		\caption {{#3}}
		\label {f:#4}
	\end {figure}
}

% Deflists:
\newlength{\defitemindent} \setlength{\defitemindent}{.25in}
\newcommand{\deflabel}[1]{\hspace{\defitemindent}\bf #1\hfill}
\newenvironment{deflist}[1]%
	{\begin{list}{}
		{\itemsep=10pt \parsep=5pt \topsep=0pt \parskip=10pt
		\settowidth{\labelwidth}{\hspace{\defitemindent}\bf #1}%
		\setlength{\leftmargin}{\labelwidth}%
		\addtolength{\leftmargin}{\labelsep}%
		\renewcommand{\makelabel}{\deflabel}}}%
	{\end{list}}%

% Equation numbering:
\makeatletter
	\newcommand{\numbereqbysec}{
		\@addtoreset{equation}{section}
		\def\theequation{\thesection.\arabic{equation}}
		}
\makeatother

% ---------------------------------------------------------------------
% General settings

\DateYMD			% the only rational default
\def\figdir{_figs}		% redefine this locally

% =====================================================================

% --- /home/prasad/.lib/tex/vgabbr.sty ---
% =====================================================================
% Personal abbreviations			v guruprasad,sep1999
% =====================================================================
% Scientific:

\def\arcdeg{\hbox{$^\circ$}}
\newcommand{\bold}[1]{\mathbf{#1}}
\newcommand\degree{$^\circ$}
\newcommand{\Qed}{$\bold{\Box}$}
\newcommand{\order}[1]{\times 10^{#1}}
\newcommand{\Label}[1]{\ \\ \textbf{#1}}
\newcommand{\Prob}[1]{\mathrm{\mathbf{Pr}}[#1]}
\newcommand{\Expect}[1]{\mathrm{\mathbf{E}}[#1]}

% Latinora:

\newcommand\ala{\emph{a la}\xspace}
\newcommand\vs{\emph{vs.}\xspace}
\newcommand\enroute{\emph{en route}\xspace}
\newcommand\insitu{\emph{in situ}\xspace}
\newcommand\viceversa{\emph{vice versa}\xspace}
\newcommand\terrafirma{\emph{terra firma}\xspace}
\newcommand\perse{\emph{per se}\xspace}
\newcommand\adhoc{\emph{ad hoc}\xspace}
\newcommand\defacto{\emph{de facto}\xspace}
\newcommand\apriori{\emph{a priori}\xspace}
\newcommand\Apriori{\emph{A priori}\xspace}
\newcommand\aposteriori{\emph{a posteriori}\xspace}
\newcommand\nonsequitor{\emph{non sequitor}\xspace}
\newcommand\visavis{\emph{vis a vis}\xspace}
\newcommand\primafacie{\emph{prima facie}\xspace}

% http://www.liv.ac.uk/education/hd/latin.html
\newcommand\circa{\emph{c.}\xspace}
\newcommand\ibid{\emph{ibid.}\xspace}		% previous citation
\newcommand\loccit{\emph{loc.\ cit.}\xspace}	% cited in the ref
\newcommand\opcit{\emph{op.\ cit.}\xspace}	% cited in the ref
\newcommand\viz{viz{}\xspace}			% videlicet - NO STOP

% http://www.tsolv.com/schools/lghs/clubs/latin/Latin_Abbreviations.html
\newcommand\ie{i.e.{}\xspace}			% id est
\newcommand\eg{e.g.{}\xspace}			% exempli gratia
\newcommand\etal{\emph{et al.}\xspace}		% et alii, et alibi
\newcommand\cf{cf.{}\xspace}			% confer (compare)
\newcommand\etc{etc.{}\xspace}			% et cetera

% General readability:
\newcommand{\figref}[1]{Fig.\ (\ref{f:#1})}
\newcommand{\figsref}[2]{Figs.\ (\ref{f:#1}-\ref{f:#2})}

\newcommand{\LD}{\begin{description}}
\newcommand{\DE}{\end{description}}
\newcommand{\LI}{\begin{itemize}}
\newcommand{\LE}{\end{itemize}}
\newcommand{\LN}{\begin{enumerate}}
\newcommand{\NE}{\end{enumerate}}
\newcommand{\VB}{\begin{verbatim}}
\newcommand{\VE}{\end{verbatim}\\}
\newcommand{\QB}{\begin {quotation}}
\newcommand{\QE}{\end {quotation}}

\newcommand{\Or}{\vee}
\newcommand{\Def}{\stackrel{\triangle}{=}}

% Documentation:
\def\ednote#1{\noindent== \emph{#1} ==}		% Editorial notes
\def\note#1{\noindent====\\\emph{#1}\\====}	% Yes notes!
\def\revising{\ednote{to be revised}}
\def\comment#1{}			% no comments!

% =====================================================================

% --- title.tex ---
\title {
	Entanglement, thermalisation and stationarity:\\
	The computational foundations of quantum mechanics
	}

\author {V Guruprasad}
\email {prasad@watson.ibm.com}
\affiliation{IBM T J Watson Research Center, Yorktown Heights, NY 10598}

%\date{2000.3.16}
%\date{1999.11.17}
\begin {abstract}
% --- abs.tex ---
'Tis said,
to know others is to be learned, to know oneself, wise -
I demonstrate that
it could be more fundamental than knowing the rest of nature,
by applying
classical computational principles and engineering hindsight
to derive and explain
quantum entanglement, state space formalism and
the statistical nature of quantum mechanics.
I show that
an entangled photon pair is literally
no more than a 1-bit hologram,
that the quantum state formalism is completely derivable
from general considerations of representation
of physical information, and that both
the probabilistic aspects of quantum theory and
the constancy of $h$
are exactly predicted by the thermodynamics of representation,
without precluding a fundamental, relative difference in spatial scale
between non-colocated observers,
leading to logical foundations of relativity and cosmology
that show the current thinking in that field
to be simplistic and erroneous.

\end {abstract}
% --- pacs.tex ---
%\pacs {
%	03.65.Bz	% Foundations, theory of measurement, misc
%	03.50.De	% Classical electromagnetism, Maxwell equations
%	05.20.-y	% Classical statistical mechanics
%	05.45.Pq	% Numerical simulation of chaotic models
%	03.67.-a	% Quantum information
%	05.30.-d	% Quantum statistical mechanics
%}
\maketitle
% --- body.tex ---
\newcommand\bra[1]{\langle#1|}
\newcommand\ket[1]{|#1\rangle}
\newcommand\iprod[2]{\langle #1 | #2 \rangle}
\newcommand\tuple[2]{\langle #1, #2 \rangle}
\newcommand\up{\uparrow}
\newcommand\dn{\downarrow}
\newcommand\definedas{\stackrel{\triangle}{=}}
\newcommand\Trace{\mathrm{Tr}}

\newcommand\Poly[1]{\bold{Poly}(#1)}
% ----------------------------------------------------------------------

\section {Motivation and overview} 
\label {s:intro}

Everyone knows that
``classical theory is \emph{absolutely incapable} of describing
the distribution of light from a blackbody''
\cite [I-41-2] {Feynman}
and just about every other result of physics,
but the opinion was formed out of frustration a century ago
when we did not have an understanding of classical mechanics itself
as we do today.
I show below how the notions of
holographic imaging and computational irreversibility suffice for
an elegant classical interpretation of
entanglement and the quantum state space formalism,
respectively, and conversely,
that the quantum postulates have been largely
a substitute for this insight.
More particularly,
the present theory is reversal of the traditional assumption
that our knowledge of physics is more fundamental than
the science of knowledge itself.

I consider
the computational issue of physical role of
the \emph{observer's physical data states}
in the process of observation.
By Landauer's principle,
these must be irreversibly altered at every observation,
unlike in Heisenberg's theory,
where the concern was only with possible disturbance to
the \emph{observed} entity.
I shall also show how
quantum uncertainty also results from this perspective,
providing a precise computational notion of smallness.
The states bear
representational correspondence to quantum wavefunctions,
yielding, as I shall show,
Schr\"odinger's equation as a computational result, and
the constancy of $h$ as a principle of scale
in communication between observers.

Of particular concern, naturally, is
the probabilistic nature of quantum information,
which inspired Schr\"odinger's cat and many-worlds interpretations,
but which, I hope to show with reasonable conviction,
is essentially thermal, and in that sense classical.
Formal notions of information have always been based on probability
arising from thermal motion,
beginning with Boltzmann's work in the kinetic theory,
but the mere attribution of classical thermalisation
does not mean that
the underlying \emph{mechanics} should be classical:
indeed, the Fermi-Pasta-Ulam (FPU) problem,
the difficulty of modelling thermal diffusion
by deterministic simulation, already casts doubt on
the adequacy of classical mechanisms to guarantee thermalisation.
On the other hand,
the notion of irreversibility in Landauer's theory
plays a key role, as I shall show,
in accounting for the probabilistic aspects of quantum theory.

Although I do not attempt to dismiss quantum mechanisms \perse,
for example particle creation and annihilation,
the result implies that
quantum effects must be indistinguishable from those
that would result from
thermodynamics and computational considerations anyway.
In the one instance
where particulate views are weakest,
\viz radiation in vaccuum,
the usual attribution of thermalisation to nonlinear interactions
becomes logically inapplicable, bypassing the FPU issue.
I have shown recently
\cite {Prasad2000a}
that classical considerations of
dynamic stationarity under wall jitter,
needed to account for the thermalisation classically,
exactly reproduce both Planck's law and the harmonic oscillators,
which forces me to consider applying classical electromagnetism
in other quantum situations as well.

As entanglement is specially of interest in quantum computation,
I demonstrate in \S\ref{s:epr} that
it is indeed reproduced by classical electromagnetic waves;
this is distinct from \emph{all} prior considerations,
including those of Bohm
\cite {Bohm1951}
\cite {Bohm1952}, and
Einstein, Podalsky and Rosen (EPR)
\cite {EPR1935},
as they invariably assumed particles in the classical picture,
and were forced into issues of
superluminal communication and hidden variables.
No such problems occur in the wave picture,
classical or otherwise,
showing that it is the particulate perception,
not classical mechanics, which has been at fault.
The conclusion is formally supported,
as discussed in \S\ref{s:states},
by a mathematical result by BenDaniel,
that representable physical information is necessarily 
one of continuous fields and not particles.
I further demonstrate that
the nature of quantum information is fundamentally holographic,
which makes entanglement purely
an artifact of \aposteriori selection,
consistent with a recent proof by Bennett \etal
of the impossibility of communicating
classical information through entanglement
\cite {BSSTT1999}.

I then establish my main contention, in \S\ref{s:states}, that
the quantum state space is itself fundamentally computational
in origin,
by reproducing its complex vector space form purely from
considerations of the most general representation for information
of any kind, as well as Schr\"odinger's equation,
generalising Dirac's derivation to any such space.
This does not suffice to prove the dependence on $h$,
for which I revert, in \S\ref{s:cavity},
to the principles of
stationarity and antinodal equipartition already established for 
the cavity spectrum, and
apply these to the classical wave formalism,
equivalent to the de Broglie waves of matter,
obtainable from Hamilton-Jacobi theory.

Lastly, I show
how these principles readily yield
the quantisation of matter, quantum uncertainty, and
the constancy of $h$,
the last as a transitive result of pairwise interactions
among observers and observed entities.
In particular,
the theory suffices to prove that
$h$ is a scale factor relating only
the dimensions of energy (mass) [M] and time [T], 
but not of space [L],
permitting a fundamental, relative difference of spatial scale
in interactions between separated entities
(\S\ref{s:summary}),
which directly leads to the logical foundations of relativity and
shows the current notions of general relativity and
cosmology to be too simplistic to have been correct
\cite {Prasad2000c}.

\section {Entanglement: 1-bit holography}
\label {s:epr}

Accordingly,
I shall first show that
classical electromagnetic waves do exactly reproduce
quantum entanglement.
This generally concerns a bipartite state of two particles of the form
$\ket{\psi^{\pm}} = \ket{a_1 b_2} \pm \ket{b_1 a_2}$,
where the subscripts denote the particles and
$a$ and $b$, their measured properties.
Entanglement lies in the inequality of
the amplitudes $\iprod{x_1 y_2}{\psi^{\pm}}$,
which are probabilistic in quantum theory,
but would be deterministic in the classical picture.
Key to our result is the observation
that the combined bras $\bra{x_i y_j}$ can be identified as
the resulting combined stationary physical state of two detectors,
to be explained in terms of computational principles, and
that this state is really
a spectral component of the overall physical state,
the temporal factor $e^{-i \omega t}$
being implicit in the notation.
Also omitted is the fact
that the ``particles'' are travelling;
the corresponding factor $e^{ikz}$ is crucial
to our classical wave interpretation.

For simplicity of argument,
I shall restrict myself to
a single Einstein-Podalsky-Rosen (EPR) experiment,
in which 
circularly polarised $\gamma$ photons are emitted by
a disintegrating source and
the photons are detected by analysing their linear polarisations.
This entails no loss of generality other than that already outlined,
\viz that the treatment is applicable only to photons,
as the resulting arguments will be extensible
even to the de Broglie waves of matter.
We denote the detection events as $\bra{x_1}$, \etc,
$x$ identifying the linear polarisation and
the subscript, the detector; and
the initial photon states as $\ket{r_1}$, $\ket{l_2}$, \etc,
representing the right- and left-circular polarisations,
respectively,
so that the entangled photon state becomes
$\ket{\psi^{\pm}}
= \ket{r_1 r_2} \pm \ket{l_1 l_2}
\equiv \ket{r_1 r_2 \pm l_1 l_2}$,
according to the parity of the parent particle.
We would describe the circular polarisation classically by
an electric vector field propagating in the $z$-direction,
\begin {equation} \label {e:ewave}
	\bold{E} (z, t)
	=
	\bold{E}_x \cos (k z - \omega t + \alpha)
	+ 
	\bold{E}_y \sin (k z - \omega t + \alpha)
	,
\end {equation}
$\bold{E}_x$ and $\bold{E}_y$ being
the amplitude component in the corresponding directions, so that
a detection at $z_i$, $i \in \{1,2\}$,
means simply that the phase, and therefore the delay $t_i$,
is determined at $z_i$,
\begin {equation} \label {e:images}
	\omega t_i
	=
		\alpha + k z_i -
	\left\{
		\begin {array}{l}
			(n + 1/2) \pi \\
			n \pi
		\end {array}
	\right.
	,
\quad
	\alpha \in [0, \pi/4)
	,
\end {equation}
according to whether
the detected polarisation was $x$ or $y$, respectively,
\ie \emph{%
	the measured bit equivalently determines 
	whether the detector was within
	an even or odd half wavelength interval from the source.%
}
With one detector, this is all we would have,
but with two non-colocated detectors,
we could associate the two detections with a common source
if, and only if, the ranges were to coincide, or in other words,
the two values represented the \emph{same bit} of information.
We can reinforce this conclusion by interpreting
the quantum kets also classically as
\begin {equation} \label {e:classikets}
\begin {split}
	\ket{r}
	\equiv
		\frac{1}{\sqrt 2}
		[ \ket{x} + \ket{y} ]
	&\approx
		\frac{1}{\sqrt 2}
		[ \bold{E}_x e^{ikz} + i \bold{E}_y e^{ikz} ]
\\
\text {and }
	\ket{l}
	\equiv
		\frac{1}{\sqrt 2}
		[ \ket{x} - \ket{y} ]
		\frac{1}{\sqrt 2}\ket{x}
	&\approx
		\frac{1}{\sqrt 2}
		[ \bold{E}_x e^{ikz} - i \bold{E}_y e^{ikz} ]
	,
\end {split}
\end {equation}
where $\approx$ denotes the correspondence and
$e^{-i\omega t}$ is once again omitted for brevity.
The analysed polarisations $\bra{x}$, $\bra{y}$
likewise correspond to the classical phasors
$\bold{E}_x^{*} e^{-ikz}$ and $\bold{E}_y^{*} e^{-ikz}$,
respectively, and lead to the classical dot products
\begin {equation}
\begin {split}
	\iprod {x_1 y_2}{r_1 r_2}
	&\approx
	\int_{z_1, z_2}
		\bold{E}_x^{*} e^{-ikz_1}
		\bold{E}_y^{*} e^{-ikz_2}
	\cdot
		\frac{1}{\sqrt 2}
		[ \bold{E}_x e^{ikz_1} + i \bold{E}_y e^{ikz_1} ]
		\frac{1}{\sqrt 2}
		[ \bold{E}_x e^{ikz_2} + i \bold{E}_y e^{ikz_2} ]
	\; dz_1 dz_2
\\
	&=
		\frac{1}{\sqrt 2}
	\int_{z_1}
		\bold{E}_x^{*} e^{-ikz_1}
		\bold{E}_x e^{ikz_1}
	\; dz_1
	\cdot
		\frac{i}{\sqrt 2}
	\int_{z_2}
		\bold{E}_y^{*} e^{-ikz_2}
		\bold{E}_y e^{ikz_2}
	\; dz_2
\\
	&= i E^2 / 2
	, \quad
	E \equiv |\bold{E_x}| = |\bold{E_y}|
\\
	\iprod {x_1 x_2}{r_1 r_2}
	&\approx
	\int_{z_1, z_2}
		\bold{E}_x^{*} e^{-ikz_1}
		\bold{E}_x^{*} e^{-ikz_2}
	\cdot
		\frac{1}{\sqrt 2}
		[ \bold{E}_x e^{ikz_1} + i \bold{E}_y e^{ikz_1} ]
		\frac{1}{\sqrt 2}
		[ \bold{E}_x e^{ikz_2} + i \bold{E}_y e^{ikz_2} ]
	\; dz_1 dz_2
\\
	&=
		\frac{1}{\sqrt 2}
	\int_{z_1}
		\bold{E}_x^{*} e^{-ikz_1}
		\bold{E}_x e^{ikz_1}
	\; dz_1
	\cdot
		\frac{1}{\sqrt 2}
	\int_{z_2}
		\bold{E}_x^{*} e^{-ikz_2}
		\bold{E}_x e^{ikz_2}
	\; dz_2
\\
	&= E^2 / 2
	,
\end {split}
\end {equation}
since terms involving both $z_1$ and $z_2$ in the exponent
vanish in the C\'esaro sum exactly as in quantum theory.
Similarly,
\begin {equation}
\begin {split}
	\iprod {x_1 y_2}{l_1 l_2}
	&\approx
	\int_{z_1, z_2}
		\bold{E}_x^{*} e^{-ikz_1}
		\bold{E}_y^{*} e^{-ikz_2}
		\frac{1}{\sqrt 2}
		[ \bold{E}_x e^{ikz_1} - i \bold{E}_y e^{ikz_1} ]
		\frac{1}{\sqrt 2}
		[ \bold{E}_x e^{ikz_2} - i \bold{E}_y e^{ikz_2} ]
	\; dz_1 dz_2
\\
	&=
		\frac{1}{\sqrt 2}
	\int_{z_1}
		\bold{E}_x^{*} e^{-ikz_1}
		\bold{E}_x e^{ikz_1}
	\; dz_1
	\cdot
		\frac{-i}{\sqrt 2}
	\int_{z_2}
		\bold{E}_y^{*} e^{-ikz_2}
		\bold{E}_y e^{ikz_2}
	\; dz_2
\\
	&= - i E^2 / 2
	,
\\
	\iprod {x_1 x_2}{l_1 l_2}
	&\approx
	\int_{z_1, z_2}
		\bold{E}_x^{*} e^{-ikz_1}
		\bold{E}_x^{*} e^{-ikz_2}
		\frac{1}{\sqrt 2}
		[ \bold{E}_x e^{ikz_1} - i \bold{E}_y e^{ikz_1} ]
		\frac{1}{\sqrt 2}
		[ \bold{E}_x e^{ikz_2} - i \bold{E}_y e^{ikz_2} ]
	\; dz_1 dz_2
\\
	&=
		\frac{1}{\sqrt 2}
	\int_{z_1}
		\bold{E}_x^{*} e^{-ikz_1}
		\bold{E}_x e^{ikz_1}
	\; dz_1
	\cdot
		\frac{1}{\sqrt 2}
	\int_{z_2}
		\bold{E}_x^{*} e^{-ikz_2}
		\bold{E}_x e^{ikz_2}
	\; dz_2
\\
	&= E^2 / 2
	,
\end {split}
\end {equation}
yielding the same result as the quantum amplitudes
(\cf \cite [III-18-3] {Feynman}),
\begin {equation} \label {e:classiamp}
	\iprod{x_1 y_2}{\psi^{\pm}}
	\approx
	\left\{
		\begin{array}{l}
			0 \\
			i E^2
		\end{array}
	\right.
\text { and }
	\iprod{x_1 x_2}{\psi^{\pm}}
	\approx
	\left\{
		\begin{array}{l}
			E^2 \\
			0
		\end{array}
	\right.
	,
\end {equation}
the difference being only that
the classical amplitudes on the right are exact and
not probabilistic.
The mystery of entanglement clearly cannot lie in
in the quantumness of the waves,
but only in their treatment as particles
\cite {EPR1935},
which introduces the notion of interaction and communication
in the first place.
The particulate perception also misses
our information argument altogether,
\viz that entanglement must signify that
the detected events concern the same bit of information,
representing a common source (eq.\ \ref{e:images}).

This form of information is well understood in holography,
where the image is formed by coincidence of antinodes
of multiple waves.
In ordinary holography,
a single wavelength is used for each colour in the image,
and angular diversity, \ie waves from multiple angles,
is used to eliminate aliases in the image by spatial coincidence.
One form of holographic radar I worked on
\cite {Prasad1986}
employed frequency diversity for the dealiasing:
we would have exactly the same principle in our EPR experiment
if our source produced photon pairs at multiple frequencies,
as eq.\ (\ref{e:images}) would become
\begin {equation} \label {e:imaging}
	\omega_j t_{ji}
	=
		\alpha + k_j z_i -
	\left\{
		\begin {array}{l}
			(n + 1/2) \pi \\
			n \pi
		\end {array}
	\right.
	,
\quad
	j = 1, 2, ...
	.
\end {equation}
With frequency-selective detectors,
we would have a pair of detected event bits
for each frequency $j$, but the aliases
belonging to any pair of frequencies $\omega_1$ and $\omega_2$
can only coincide within intervals corresponding
to the larger of the two wavelengths
\{$\lambda_1 \equiv c/\omega_1$, $\lambda_2 \equiv c/\omega_2$\},
thus diminishing the density of aliases,
as well as refining the uncertainty interval to the smaller wavelength.
Even with a limited number of query frequencies,
it is thus possible to localise the source
within a known neighbourhood.
In the EPR case,
we have only one bit of coincidence information, and thus
an infinite number of equally significant aliases,
which made the information hard to recognise.

We still need to reexamine the issue of superluminal communication,
which implies the conjecture that
the detected value at $z_2$ can be influenced
by a measurement at $z_1$ \emph{after}
the emission of the photons from the source.
A stronger result,
that entanglement as such confers no ability whatsoever
to communicate classical information,
has in any case already been established recently
from within the confines of quantum theory
\cite {BSSTT1999},
but we need to be able to arrive at the same result
within our imaging interpretation.
One difficulty,
which could have made our classical wave analysis
unthinkable in the past, is
that eq.\ (\ref{e:ewave}) describes a pure, endless sinusoidal wave,
not a wave packet with which we could associate
a discrete detection event and time-of-flight issues.
Even the quantum wavefunctions in the EPR scenario, however,
do not denote wave packets at all,
but merely individual spectral components,
\ie pure sinusoids, and moreover,
we have already established that
the multiple spectral components forming a wave packet
would only yield a sharper image.
Accordingly,
we only need to verify that
communication \perse does not occur in
the present interpretation either, and
the reason may be recalled from past theory concerning relativity,
\viz
that the information represented by the measurement at $z_2$
is available only \emph{after} correlation with $z_1$,
signifying \aposteriori identification.

The EPR correlation is thus mysterious only in the particulate view,
which must hence be wrong.
Further,
although I showed entanglement
to be classical only in the case of photons,
the imaging principle is identically applicable to any wave formalism,
including quantum wavefunctions, meaning that in all such cases,
entanglement simply concerns independent detections of a common source.
Two other difficulties exist in this regard,
first,
	that a similar classical wave analysis
	does not appear possible for particles in general,
and second,
	that the quantum wavefunctions and amplitudes are probabilistic,
	which is again hard to reconcile with
	the classical mechanics of particles.
These are the very ones that make
the wave-particle duality conceptually difficult
in the first place, as the wave aspect by itself
is inherently common to both classical and quantum descriptions,
as evidenced above.

Accordingly,
I show in the remaining sections that
these difficulties too vanish on applying
other modern engineering principles
from control, computation and thermodynamic theories,
the principle in each case
being based solely on precise classical reasoning.
In \S\ref{s:states},
I shall first show that
the quantum formalism of states is in fact simply 
the most general representation of information, and therefore
the representation necessary and sufficient
for analysing the most complex situations
including particle physics and the computational principles
of the brain.
I shall then show, in \S\ref{s:cavity},
how this formalism gets represented and linked to observed physics,
in particular,
how we arrive at a nonzero $h$ and its constancy, as well as
at the probabilistic character and the uncertainty principle
of quantum mechanics,
also on basis of sound classical, but modern, engineering.

\section {Representability: the anti-particulate principle}
\label {s:states}

The contention, essentially, is that 
the quantum formalism of states is foremost computational, and
then acquires its probabilistic character
once again because of
a fundamental principle of computation related to
the thermodynamics of representation.
The separation was unobvious in the past because 
the probabilistic nature suggests a closer relation to Shannon's theory,
obfuscating the representational aspect of
the observer's data states, and
yet not revealing the mechanisms
that cause it to be statistical in the first place.
The distinction is subtle but fundamental,
as the logical notion of representation is \perse deterministic,
despite the fact that
any physical embodiment is bound to be subject to thermal erosion.

Representability, or definability, as such appears to be 
sufficient to imply inherent quantisation of fields,
as recently shown by BenDaniel
\cite {BenDaniel1999},
the equivalent computational argument being that
the total knowledge of any finite set of observers
can only be finite to the extent
it is representable and communicable, using 
a finite number of sentences in a language with a finite alphabet,
as any other knowledge would be inexpressible by definition and
therefore outside the realm of science.
Evidently,
the premise could have been used avoid many of
the measure-theoretic difficulties
historically encountered in quantum theory,
as suggested in Appendix \ref{a:calculus},
where Cauchy's notion of continuity is also shown to be
literally equivalent to our consideration of finite representation.
More particularly implied is that:

\emph{%
Physical information fundamentally represents continuous fields
and this it can do perfectly; conversely,
particles can never be perfectly represented or known,
nor directly represented.%
}

One may recall that in classical mechanics,
a particle is considered to be a geometrical point
requiring no spatial description beyond
a triplet of real numbers defining its instantaneous location,
with possibly a second triple specifying its velocity.
The distinction is that
the point coordinates do not suffice to \emph{delimit} the object
as a point particle;
to delimit without contextual knowledge requires
infinite bits of representation.
For example, as already explained,
with many bits of coincidence information,
we can locate a source very closely,
albeit still with infinite alias regions,
so that contextual knowledge is needed to limit
the localisation to one neighbourhood.
With infinite information we would be able to locate
the source \emph{exactly} and with no contextual knowledge at all,
but the exercise assumes that our source is a point entity.
No such difficulty occurs
if we accept our representative data as
inherently referring to a continuous field and not a point particle.
This principle formalises our contention in \S\ref{s:epr} that
the particulate view is as such erroneous and
partly responsible for
the semantic difficulties of quantum theory,
as will be illustrated again with respect to
the notion of intrinsic spin.

To demonstrate the computational origin of the state space formalism,
we now attempt to construct the formalism purely by seeking
the most general representation for knowledge of any kind. 
We could pick arbitrary variables, for instance, 
a combination of groceries, train schedules and stock prices.
Despite the generality,
we would still have two conditions
available in any such choice, \viz
	numerical values and,
	if pertaining to real phenomena,
	observability at various times.
In the resulting multidimensional state space,
the observed data would be essentially represented
in the direction of state vectors,
in absence of a meaningful norm for all such spaces.
Differences between states could be trivially treated as vectors, and
the temporal evolution of a state would be represented
only as a continuing change of direction.
As shown in Appendix \ref{a:calculus},
we do not need to be able to measure the represented variables
with infinite precision at infinitely small intervals;
rather, it is our inability in both regards
that limits us to finite data of finite precision,
and therefore to continuous fields,
representing the impossibility of point delimitation,
as just explained.

The quantum equations of motion are then obtained
as general computational result,
by considering the most general form of temporal evolution
over such a space.
By the representabilty principle,
this too must be limited to
a finite number of coefficients of finite precision,
the latter being implicit, yielding
\begin {equation} \label {e:nevolution}
	\sum_{i = 0}^{n}
		a_n \frac{d^n}{dt^n} \ket{\psi}
	=
		\ket{\phi}
	=
		F^{(n)} (t) \ket{\psi}
	,
\end {equation}
where the kets merely denote our arbitarily chosen state space, and
$F^{(n)}(t)$ signifies
a continuously varying applied ``force'' on the system,
assuming our variables are sufficiently well behaved.
The generality is the reason we may apply techniques
from Laplace transform theory to solve eq.\ (\ref{e:nevolution});
in particular,
the operator sum $\sum_{i=0}^{n} a_n d^n/dt^n$
transforms to $\sum_{i=0}^{n} a_n s^n$, and
yields the characteristic equation $\sum_{i=0}^{n} a_n s^n = 0$,
so that by the fundamental theorem of algebra,
the complex plane becomes necessary and sufficient
to represent all possible patterns of evolution.
We have thus reproduced the broad notion of state space
as well as the need for complex valued representation,
but without \apriori assumption of quantisation,
so that these ideas are valid for any state space,
not just one of quantum theory.

The general state space notion of evolution
automatically leads to Schr\"odinger's equation, as shown by Dirac
\cite [\S27] {Dirac},
because the derivative of the incremental evolution
$\ket{\psi(t)} \rightarrow \ket{\psi(t+ \Delta t)}$
must have the operator form
\begin {equation} \label {e:se}
	i \hbar \frac{d}{dt} \ket{\psi} = H (t) \ket{\psi} ,
\end {equation}
where the imaginary coefficient $i$ derives from the fact
that vector magnitudes lose significance in state representation, and
$\hbar$ is merely a scale factor relating
the rate of evolution operator $H$ to the observer's scale of time.
This also leads to spectral decomposition in quantum theory
\cite [\S29] {Dirac},
but it should be realised that in general,
$e^{i \omega t}$ is simply the imaginary component of
(the eigenfunctions of) the Laplace operator $s \sim d/dt$.
It is only by restricting to stationary states that
the real parts, representing transients $\propto e^{- \alpha t}$,
vanish from quantum theory, and
the reasons both for stationarity and the constancy of $\hbar$
in eq.\ (\ref{e:se})
are also basically computational,
as will be explained in \S\ref{s:cavity}.

The representational freedom explains
why we were able to drop $e^{ikz}$ in
the quantum versions of eq.\ (\ref{e:classikets}) and
yet arrive at the entanglement prediction, eq.\ (\ref{e:classiamp});
we have paid for our semantic imprecision, of course,
by confusing the result with superluminal communication.
The quantum notion of spin,
as a mysteriously intrinsic property of particles,
results from a similar imprecision:
if in light of the foregoing theory, we recognise
a particle wavefunction
$\ket{\psi} \equiv
\left( \begin{array}{l} \psi_x \\ \psi_y \end{array} \right)$
as a continuous physical field, and remember
that we are referring to a travelling wave component,
so that both spatial and temporal factors
$e^{ikz}$ and $e^{i\omega t}$, respectively,
are required for its complete physical representation,
it becomes clear that the effect of a Pauli spin matrix,
say $\left( \begin{array}{ll} 0 & -i \\ i & 0 \end{array} \right)$,
on $\ket{\psi}$ is to simply rotate
the local direction of motion as one travels within the field,
literally describing a vortex field rather than
an inherently non-geometrical form of angular momentum.
It is only in the particulate view,
where the complex exponential factors get dropped,
that we end up with representational, and consequently semantic, loss.
On a related note, in ignoring the magnitudes,
we had also lost the $E^2$ in eq.\ (\ref{e:classiamp}), and
this had cost us insight from classical electromagnetism.

\section {Thermodynamic stationarity}
\label {s:cavity}

I still need to show
how $\hbar$ acquires
the special significance it does in quantum theory,
when applied to mechanical information, and
account for the probabilistic nature of quantum wavefunctions,
else the foregoing theory would be considered
no more than a coincidence.

The first part has been recently achieved
\cite {Prasad2000a}
by going back to Planck's theory, and proving that thermal wall jitter,
which is in fact necessary to account for
the thermalisation of radiation in the first place
as there are no nonlinear interactions \emph{between} photons
that might suffice for this purpose,
unlike the case of gas molecules in kinetic theory;
is in fact sufficient to exactly reproduce
both Planck's postulates and the radiation spectrum law.
The wall jitter continually changes the stationary modes of the cavity,
not just the energy distribution therein;
we need to consider only the fraction of thermal changes
that are slower than the electromagnetic transit time 
across the cavity,
because faster motions can only affect transient behaviour.
We therefore sum over the equilibrial distribution of modes,
and in the process identify families of modes
that are related by whole numbers of antinodes.
It is trivial to verify that within such a family,
the frequencies would be necessarily related as
$\{f$, $2f$, $3f$, ...$\}$, and importantly,
incremental wall displacements can only change the modal distribution
by an integral number of antinodes,
\ie between members of the same family,
so that, discounting transients,
the harmonic families behave exactly as Planckian oscillators,
as each family can have exactly one member energised at any instant.
The energy of an antinodal lobe depends only on
the amplitude and not the frequency, and in the equilibrial state,
the antinodal lobes must have the same energy,
reproducing Planck's quantisation rule
\begin {equation} \label {e:qrule}
E = h f ;
\end {equation}
the sums over the families then trivially reproduces
the equations in Planck's derivation.
The detailed balance,
needed for balancing the various contributions,
from the thermal Doppler spread due to wall jitter,
possible interactions with the wall atoms,
nonlinearities due to imperfect vacuum within the cavity,
and so on,
turns out to be identical in form to the Bose-Einstein derivation.
Planck's law
\begin {equation} \label {e:Planck}
	\widehat{U} (f)
	= \frac{ hf }{ e^{hf/k_B T} - 1 } ,
\end {equation}
$\widehat{U} (f)$ denoting
the equilibrial expection energy at frequency $f$,
thus turns out to be a strictly classical result,
stemming from classical electromagnetics and classical thermalisation.

More importantly,
the very form of eq.\ (\ref{e:Planck}) shows
that $h$ relates to the antinodal lobes and the spectral domain
in almost the same way as the Boltzmann constant $k_B$ does to
gas molecules and the spatial domain,
as it ``exposes'' $h$ to measurement the same way as
the law of Brownian motion
$\langle R^2 \rangle = 6 k_B T t / \mu$,
$R$ being the mean radial distance travelled
as a function of time $t$, and $\mu$, the mobility;
historically exposed $k_B$,
allowing it to be empirically determined from diffusion phenomena
\cite [I-41-4] {Feynman}.
It should also be noticed that
an antinodal lobe is indeed
the $\lambda/2$ interval of uncertainty
we encountered when interpreting entanglement (\S\ref{s:epr}),
so that
our holographic notion of information would be consistent with
a stationary representation in thermodynamic equilibrium,
to be explored via Hamilton-Jacobi formalism below. 

The constancy of $h$ was also independently established
using Dirac's result that
given any anti-commuting relation $[.,.]$ and
two pairs of conjugate variables
$u_1$, $v_1$ and $u_2$, $v_2$, we would get,
with no assumption of dependence between the pairs,
the cross-relation
\begin {equation}
	[ u_1 , v_1 ] (u_2 v_2 - v_2 u_2) = 
		(u_1 v_1 - v_1 u_1) [ u_2 , v_2 ]
\end {equation}
implying that they must have
the same constant of anticommutation
\cite [\S21] {Dirac},
\begin {equation} \label {e:Commute}
\begin {split}
	u_1 v_1 - v_1 u_1 &= K [ u_1 , v_1 ]
\\
\text {and }
	u_2 v_2 - v_2 u_2 &= K [ u_2 , v_2 ] .
\end {split}
\end {equation}
We could, for instance,
choose classical electric and magnetic wave amplitudes
$\bold{E}$ and $\bold{B}$ within our cavity
as $u_1$ and $v_1$, and
the induced emf $V$ and current $i$ in a probe
for $u_2$ and $v_2$.
We should then get the same value for $K$
for the anticommutation of $V$ and $i$ as for $E$ and $B$.
We could next reuse the symbols $u_1$ and $v_1$ for $V$ and $i$,
respectively, and identify $u_2$ and $v_2$ with
the dynamical variables of a third system,
establishing the validity of $K$ for the third system,
and so on, proving that
every physical system
that could directly or indirectly interact with our cavity
would be similarly quantised with the same value of $h$,
proving it to be a universal constant by
transitivity of interaction.

For a general notion of waves and their stationarity
in the context of matter, we must, as previously mentioned,
turn to Hamilton-Jacobi theory,
beginning with the following relation
between Hamilton's principal and characteristic functions,
$S$ and $W$ respectively,
for a single particle system in Cartesian coordinates
\cite [\S10-8] {Goldstein},
\begin {equation} \label {e:HJwave}
	S (q,P,t) = W (q,P) - E t ,
\end {equation}
$q$ being the particle's generalised coordinate,
$E$, its total energy, and $P$, its momentum,
which is a constant of the motion in the configuration space
for which $S$ is a solution of the Hamilton-Jacobi equation
\begin {equation} \label {e:HJ}
	H (	q_1, ... , q_n ;
		\frac{\partial F_2}{\partial q_1}, ...,
		\frac{\partial F_2}{\partial q_n}
		; t )
	+
		\frac{\partial F_2}{\partial t}
	=
		0
	.
\end {equation}
$W$ closely relates to
the action-angle variables used in the early days of quantum mechanics,
the angle variable $w$ being defined as $\partial W / \partial J$,
$J$ being the action variable:
one would first solve the classical problem using these variables, and
quantisation was achieved by replacing $J$ with
a multiple of $h$.
The approach dropped out of vogue for two reasons,
first,
	the classical model was not always solvable and
	the state space formulation proved simpler and more general,
	for reasons we have already established,
and second,
	the cause of quantisation was then not known,
	so that one could not aspire for deeper understanding
	by sticking to the classical formalism.
The second reason is now no longer valid,
as the cause of quantisation has been uncovered
in the context of radiation equilibrium.
It is meaningful, therefore,
to reexamine the classical approach in this light,
particularly to seek a general classical formalism
of travelling waves to which
we could apply our antinodal holographic ideas.

Eq.\ (\ref{e:HJwave}) describes constant-$S$ wavefronts
travelling in the same direction as the particle:
the particle's velocity is given by $u = E/p = E/mv$, and
we also have $\bold{p} = \nabla W$,
but the ``waves'' are not given to be periodic.
We need a spectral decomposition, and since,
as Goldstein points out,
$S = W - Et$ must be proportional to the phase $\omega t$
and $W$ is independent of time,
$E$ must be proportional to $\omega$.
Since this is similar to
Planck's quantisation rule (eq.\ \ref{e:qrule}),
it was traditionally assumed that
classical mechanics stopped just short of quantum theory,
as it gave no reason for assuming the constant of proportionality
to be nonzero.
Indeed,
a plane wave trial solution of the form
$\psi = \psi_0 e^{iS/\hbar}$,
corresponding to $E = \hbar \omega$,
would turn Schr\"odinger's equation (\ref{e:se}) to
\begin {equation} \label {e:HJS}
	\left[
		\frac{1}{2m}
		(\nabla S)^2 + V
	\right]
		+
		\frac{\partial S}{\partial t}
	=
		\frac{i \hbar}{2m} \nabla^2 S
	,
\end {equation}
via the following well-known form of eq.\ (\ref{e:se})
\begin {equation}
	\frac{\hbar^2}{2m} \nabla^2 \psi
	-
		V \psi
	=
	\frac{\hbar}{i} \frac{\partial \psi}{\partial t}
\end {equation}
and the derivatives
\begin {equation}
	\frac{\partial \psi}{\partial t}
	=
	\frac{i}{\hbar} \psi \frac{\partial S}{\partial t}
\text { and }
	\frac{\partial \psi}{\partial x}
	=
	\frac{i}{\hbar} \psi \frac{\partial S}{\partial x}
\end {equation}
which yield
\begin {equation}
	\nabla^2 \psi
	=
	\frac{i}{\hbar} \psi \nabla^2 S
	-
	\frac{\psi}{\hbar^2} (\nabla S)^2
\end {equation}
for the Laplacian of $\psi$.
Eq.\ (\ref{e:HJS}) would reduce to
the Hamilton-Jacobi equation (\ref{e:HJ}) if $\hbar$ were to vanish,
and this was the basis for Bohr's Correspondence Principle (BCP),
that quantum mechanics reduces to classical theory
in the short-wavelength limit, as
	$\hbar \nabla^2 S \ll (\nabla S)^2$
is equivalent to, for a one-particle system,
	$(\lambda/p) \; (dp/dx) \ll 2\pi$.

Observe, however,
that for the validity of eq.\ (\ref{e:HJwave}),
we could not possibly set $E = 0 \cdot \omega$, \ie 
the BCP is not classically meaningful at all.
The error lies in mistaking
the trial solution $\psi_0 e^{iS/\hbar}$
to be a complete description,
instead of as a mere spectral component of the overall dynamics.
In the spectral decomposition,
we would look for stationary modes for which
the $Et$ term in eq.\ (\ref{e:HJwave}) vanishes anyway, and
we would not need $\hbar$ to vanish,
because the right hand term in eq.\ (\ref{e:HJS}) then
merely refers to a Fourier component,
not the whole picture.

Importantly,
\emph{stationarity is a necessary condition for both
the physical data states of the observer and of the observed entities,
as both the observer's knowledge and the represented information
cannot change between observations.}
This is an essentially computational principle,
but it forces us to consider
only the stationary states of material systems
as being physically relevant.
The stationarity further means that
the Hamiltonian becomes completely separable
(\cf \cite [\S10-6] {Goldstein}), so that
the generalised coordinates $q_i$ and momenta $p_i$
are related as in-phase and quadrature components, respectively,
at each of the characteristic frequencies.
In particular,
the combined stationary states of radiation and matter
must be stationary with respect to either, and
we can then apply Dirac's transitivity argument
(eq.\ \ref{e:Commute})
to each such pair \{$q_i, p_i$\} and
the \{$\bold{E}_i$, $\bold{B}_i$\} radiation amplitudes
at the same frequencies $f_i$, to infer that
the antinodal equipartition must be identically applicable
to these stationary modes of matter.

The stationarity and equilibrium principles fully explain
both quantisation and the probabilistic aspects of quantum theory.
As example, recall that
radiation quantisation was conceived to in order to explain
the almost instantaneous nature of photoelectric emission
\cite {ResnickHalliday}, but
by including the observer's data states in the picture,
we can readily see that the abruptness is logically unavoidable,
as the observer cannot know of states intermediate
to those representable within its own material embodiment.
The condition of thermal equilibrium,
necessary for potential long term stability of
the observer's data states, guarantees, 
via the spectral decomposition of the Hamilton-Jacobi wave analysis
of the observer's embodiment described above, that
the data states can only again change in antinodal increments
of the same energy as those of radiation.
We may therefore represent the observed variables by kets $\ket{\psi}$,
the detector values, representing the observer's state space,
by bras of the form $\bra{\phi}$, and
get the same amplitudes $\iprod{\phi}{\psi}$
as when $\bra{\phi}$ is set to represent
a second observed entity instead, interacting with the first.
In either case, the amplitude must be complex
because of the algebraic completeness of the state space formulation.
The amplitude is consistent with the (square root of the) probability
of the observed system \emph{and} the observer
\emph{thermally} and \emph{irreversibly}
transiting to the combined state $\tuple{\phi}{\psi}$,
the irreversibility being implied by the condition that
the $\bra{\phi}$ must be itself equilibrial and capable of
lasting till erasure or the next observation.
As particularly described by Landauer
\cite {Landauer1961},
the irreversibility means that
the final state is attained regardless of
the previous states of the observer and the observed,
reproducing the apparent ``collapse'' of the quantum wavefunction
in the process of observation.
Lastly, the fact that
the transitions can only occur in increments of an antinodal lobe
reproduces Heisenberg's uncertainty principle.

\section {Summary and loose ends} 
\label {s:summary}

To summarise,
I have established, though not in this order,
\LN
\def\theenumi{\roman{enumi}}
\item
	\label {i:general}
	that the quantumstate space formalism
	is no more than the most general representation
	of information of any kind whatsoever,
	which would account for its discovery
	in the first place, and
	its applicability in diverse fields
	ranging from particle physics to studies of the brain
	(\S\ref{s:states});
\item
	\label {i:fourier}
	that quantum wavefunctions merely constitute
	Fourier in-phase and quadrature components,
	as already familiar to electrical engineers
	working with strictly classical electromagnetism
	(\S\ref{s:states}, \S\ref{s:cavity}), and
	that their quantisation and randomness are
	due to separate computational and thermodynamic causes
	(\ref{i:cavity}, \ref{i:datastates}, \ref{i:constancy});
\item
	\label {i:debroglie}
	that the de Broglie waves of matter are obtainable
	from Hamilton-Jacobi theory, and
	that the Correspondence Principle was erroneously conceived
	in the context because
	the computational requirements of
	stationarity and thermal equilibrium
	(\ref{i:datastates}) were not known;
\item
	\label {i:algebra}
	that complex values occur in quantum mechanics because
	of the generality of the information represented,
	as a result of the fundamental theorem of algebra,
	which incidentally proves
	the computational completeness of the quantum formalism
	that had not been established in prior theory,
	supporting (\ref{i:general}), and
	relates Schr\"odinger's equation to control theory
	(\S\ref{s:states});
\item
	\label {i:epr}
	that quantum entanglement,
	as best illustrated by the EPR paradox,
	amounts simply to 1-bit holography in
	the classical wave picture (\S\ref{s:epr}), and
	that issues of hidden variables and
	action-at-a-distance are merely 
	consequences of preconceived particulate view;
\item
	\label {i:contextfree}
	that the particulate view in fact contradicts
	formal considerations of
	the representability of physical information,
	which show that only continuous fields can at all
	be physically represented,
	and thus observed or communicated,
	without contextual bias,
	validating
	the present notions of entanglement (\ref{i:epr}),
	Hamilton-Jacobi waves (\ref{i:debroglie}) and
	the classical field interpretation of spin
	(\S\ref{s:states});
\item
	\label {i:cavity}
	that both quantisation and
	the probabilistic nature of quantum wavefunctions
	are completely explained by the same principles of 
	stationarity and antinodal equipartition
	(\S\ref{s:cavity});
\item
	\label {i:datastates}
	that both stationarity and thermal equilibrium are
	mandated by the computational consideration of
	stability of the observer's data states and
	the represented knowledge (\S\ref{s:states}),
	connecting the representational generality
	(\ref{i:general}, \ref{i:algebra}, \ref{i:contextfree})
	with the actual physics
	(\ref{i:fourier}, \ref{i:debroglie},
	\ref{i:epr}, \ref{i:cavity});
\item
	\label {i:fluct}
	that the Hamilton-Jacobi analysis (\S\ref{s:cavity})
	establishes not only the sufficiency of
	stationarity and thermalisation
	for (\ref{i:cavity}),
	but also their necessary involvement in this role,
	obviating any postulate or alternative explanation;
and
\item
	\label {i:constancy}
	that the constancy of $h$ is purely a consequence
	of the transitivity of these constraints
	between observed systems as well as between physical observers,
	signifying the establishment of a universal scale,
	given that $h$ relates the (independent) dimensions
	of energy (mass) and time, by communication,
	again a matter of logic and information.
\NE
To illustrate the attribution of
quantum probabilities to thermalisation,
consider the hypothetical case of an observer
frozen to absolute zero temperature.
By the third law of thermodynamics (Nernst theorem),
such an observer would be frozen into one state of knowledge and
would be incapable of making any observations whatsoever.
An interesting example of the obfuscation
by notions of probability and information theory in the past is
the following observation that
the germs of BenDaniel's result were already present in kinetic theory:
Boltzmann's notion of thermodynamic information
as the spatial localisation of a gas molecule
closely parallels our notion of particle delimitation, and
the randomness of molecular motion had nothing directly to do
with this measure,
as with each bit of information serves to (deterministically)
halve the region of uncertainty.
The information and randomness aspects were thus separable in
Boltzmann's theory,
but were not recognised as such because 
the determinism properly belongs to
the computational issue of representation, and
the present distinction of computational and informational aspects
does not seem to have been at all made in prior literature.

I have not attempted to discuss how
the present theory should be applied to
other quantum issues like
superconductivity,
exchange coupling,
\emph{zitterbewegung},
particle creation and annihilation,
broken symmetries, or the grand unified theories (GUTs), \etc
These are simply too numerous to be recast or corrected
by one treatise or author.
The fundamental principles of quantum mechanics,
which are basis for all such applications,
have already been shown to be
fundamentally computational and thermodynamic.
The fact that the equilibrial antinodal lobe,
representing the quantum of change
in both radiation and matter waves as described,
is independent of the wavelength $\lambda \equiv c/f$ means that
the establishment of thermal equilibrium and
quantum mechanical consistency
do not depend on \apriori equality of spatial scale
between the interacting entities.
The hypothesis of relative spatial scale suffices for deducing both
the postulates of relativity and
Maxwell's equations of electromagnetism, and more importantly,
shows that
the current ideas in general relativity and
cosmology are entirely too simplistic,
as separately described in
\cite {Prasad2000c}.
There again,
the problem in prior theory has never been
a shortage of mathematical skills,
but of the computational intuition of
the representability of physical information, and
the cognition of fundamental limitations arising from this constraint.

\begin{acknowledgements}

To my colleagues A Joseph Hoane and Daniel Oblinger
for valuable discussions involving my EPR solution.

\end{acknowledgements}

% end
\appendix
% --- xfinite.tex ---
\section {Finite domain calculus}
\label {a:calculus}

The premise of representational finiteness allowed me, in 1983-84,
to define a notional framework as follows:
\LN
\def\theenumi{\Alph{enumi}}
\item
	Every function $f$ is represented by
	at most a finite number of symbols
	denoting algebraic variables and operations.
	The continuity of a domain $X$ is
	likewise defined as the condition
	that at any finite cardinality $\#(X)$,
	additional points may be \emph{physically} introduced,
	\emph{by resizing, adjustment of magnification,
	or technological replacement, \etc}
	to indefinitely define new points in
	the neighbourhood $N(x)$ of every point $x \in X$.

\item \label {i:cauchy}
	Continuity of a function $f: X \rightarrow Y$
	is then defined by the conditions that
	$X$ and $Y$ are both continuous,
	every new point $x' \in N_{\epsilon}(x)$
	introduces a computed value $y' = f(x') \in N_{\delta}(y)$,
	as in Cauchy's definition.
\NE
The point is that Cauchy's definition involves only
finite domains at any finite stage in the implied execution
of the limit operations, and
even the issues of divergence in quantum field theory, for instance,
are really concerned with how
computed behaviour varies with the precision of the referenced domain,
which is implicitly finite in all of mathematics.
Like the intuition of continuity given by BenDaniel's result
(\S\ref{s:states}),
this is too simple to be obvious.

% end
% === cat *.bbl ===

% --- \input.tex ---
\end {document}